\documentclass[preprint,aps,nofootinbib]{revtex4}
\usepackage{graphicx}
\input{epsf.sty}
\usepackage{epsfig}

\newcommand{\xw}{x_{W}}

\def\lsim{\mathrel{\raise.3ex\hbox{$<$\kern-.75em\lower1ex\hbox{$\sim$}}}}
\def\gsim{\mathrel{\raise.3ex\hbox{$>$\kern-.75em\lower1ex\hbox{$\sim$}}}}

\def\singleandabitspaced{\baselineskip=\normalbaselineskip\multiply
    \baselineskip by 150\divide\baselineskip by 100}

\def\singlespaced{\baselineskip=\normalbaselineskip}

\begin{document}

\singlespaced

\hfill$\vcenter{\hbox{\bf hep-ph/0601142}}$
\vskip 0.4cm

\title{TeV-Scale Stringy Signatures at the Electron-positron Collider\\
} 
\author{Piyabut Burikham \footnote{piyabut@iastate.edu}}
\vspace*{0.5cm}
\affiliation{Department of Physics and Astronomy, Iowa State University, \\
        Ames, Iowa 50011, USA} 
 
\date{\today}

\vspace*{1.0cm}

\begin{abstract}
We investigate the TeV-scale stringy signals of the four-fermion scattering at the electron-positron collider with the center of mass energy $500-1000$ GeV.   The nature of the stringy couplings leads to distinguishable asymmetries comparing to the other new physics models.  Specifically, the stringy states in the four-fermion scattering at the leading-order corrections are of spin-1 and 2 with the chiral couplings inherited from the gauge bosons identified as the zeroth-mode string states.  The angular left-right, forward-backward, center-edge asymmetries and the corresponding polarized-beam asymmetries are investigated.  The low-energy stringy corrections are compared to the ones induced by the Kaluza-Klein~(KK) gravitons.  The angular left-right asymmetry of the scattering with the final states of $u$ and $d$-type quarks, namely $c$ and $b$, shows significant deviations from the Standard Model values.  The center-edge and forward-backward asymmetries for all final-states fermions also show significant deviations from the corresponding Standard Model values.  The differences between the signatures induced by the stringy corrections and the KK gravitons are appreciable in both angular left-right and forward-backward asymmetries.
          
\end{abstract}

\maketitle


\newpage

\setcounter{page}{2}
\renewcommand{\thefootnote}{\arabic{footnote}}
\setcounter{footnote}{0}
\singleandabitspaced

\section{Introduction}

There is a number of new physics~(NP) models beyond the standard model~(SM) of particle physics.  Motivated by the hierachy and/or the fine-tuning problem in the SM, most NP models propose new states with TeV-scale masses.  A few examples are the SUSY models, models with extra gauge bosons~($Z^{\prime}$ models), and the models with extra dimensions.  When the masses of the NP states are heavier than the center of mass~(CM) energy of the collider, the effects of the NP can be measured indirectly in terms of the deviations of the SM observables such as the total cross section and various asymmetries.  The deviations from the SM in the scattering processes are determined by the mass, spin, and coupling strength of the new states being exchanged by the initial state particles.        

The question of how to distinguish new states with different spins and couplings at the low energies arises at the sub-TeV $e^{+}e^{-}$ collider.  While the CERN Large Hadron Collider~(LHC) will probe NP models with the TeV-scale masses, we certainly need the precision measurements to distinguish signature of one model from the others.  The precision measurements of the four-fermion scattering at the $e^{+}e^{-}$ collider are expected to efficiently reveal the nature of the intermediate states being exchanged by the fermions.  The angular distributions and the asymmetries induced by various new states provide information of the spin and coupling of the interactions.  

At the International Linear Collider~(ILC) with the center of mass energy $E_{cm}=500-1000$ GeV, the TeV masses could not be observed directly as the resonances since they are heavier than $E_{cm}$.  Low energy Taylor expansion is a good approximation for the signals induced from the NP models and the corrections will be characterized by the higher dimensional operators.  For the 4-fermion scattering, the leading-order NP signals from the states with spin-0 and spin-1 such as leptoquarks, sneutrino with $R$-parity violating interactions~\cite{krsz} and $Z^{\prime}$ will appear as the dimension-6 contact interaction at low energies.  As a candidate for the NP state with spin-2, the interaction induced by the (massive)~gravitons, $h^{\mu \nu}T_{\mu \nu}/M_{Pl}$, can be characterized at the low energies by the effective interaction of the form $-i\lambda T^{\mu\nu}T_{\mu\nu}/M^{4}_{H}$~\cite{hew,cpp}, a dimension-8 operator.  In the viewpoint of effective field theory, this effective interaction does not need to be originated from the exhanges of massive graviton states and it is not the most generic form of the interaction containing dimension-8 operators.  However, it certainly has the gravitational interpretation due to the use of the symmetric energy-momentum tensor $T_{\mu\nu}$.  It can be thought of as the low-energy effective interaction induced by exchange of the KK gravitons~(in ADD~\cite{add,kk} and RS~\cite{RS1} scenario) interacting with matter fields in the non-chiral fashion.  

In the braneworld scenario where the SM particles are identified with the open-string states confined to the stack of D-branes subspace, and gravitons are the closed-string states propagating freely in the bulk spacetime~\cite{stringy,tevst}, table-top experiments~\cite{gra} and astrophysical observations allow the quantum gravity scale to be as low as TeVs~\cite{add}.  Since the string scale, $M_{S}$, in this scenario is of the same order of magnitude as the quantum gravity scale, it is possible to have the string scale to be as low as a TeV.  The TeV-scale stringy excitations would appear as the string resonances~(SR) in the $2\to 2$ processes at the LHC~\cite{bfh}.  The most distinguished signals would be the resonances in the dilepton invariant mass distribution appearing at $\sqrt{\hat{s}}=\sqrt{n}M_{S}, n=1,2,...$  Each resonance would contain various spin states degenerate at the same mass.  These SRs can be understood as the stringy spin excitations of the zeroth modes which are identified with the gauge bosons in the SM.  They naturally inherit the chiral couplings of the gauge bosons.          

In this article, it will be shown that the leading-order stringy excitations of the exchanging modes identified with the gauge bosons in the four-fermion interactions will contain both spin-1 and 2.  Their couplings will be chiral, inherited from the chiral coupling of the zeroth mode~(identified with the gauge boson).  Namely, we construct the tree-level stringy amplitudes with chiral spin-2 interactions~(in addition to a stringy dimension-8 spin-1 contribution contrasting to the dimension-6 contributions from other $Z'$ model at low energies) which cannot be described by the non-chiral effective interaction of the form $-i\lambda T^{\mu\nu}T_{\mu\nu}/M^{4}_{H}$ as stated above.  

This article is organized as the following.  In Section II, we discuss briefly the construction of the stringy amplitudes in the four-fermion scattering as is introduced in the previous
work~\cite{has,cpp,cim,fhh,bhhm,bfh}, the comments on the chiral interaction are stated and emphasized.  In Section III, the low-energy stringy corrections are approximated.  The angular momentum decomposition reveals the contribution of each spin state induced at the leading order.  In Section IV, we calculate the angular left-right, forward-backward, and center-edge asymmetries.  The extensions to partially polarized beams are demonstrated in Section V.  In Section VI, we make concluding remarks and discussions.  The low-energy~($E<M$) expressions for the asymmetries $A_{LR}(z), A_{FB}, A_{CE}(z^{*})$ induced by the SM and the NP models~(KK gravitons and SR) up to the order of $O((E/M)^{8})$ are given in the Appendix.
              
\section{Open-string amplitudes for the four-fermion interactions}

  The 4-fermion processes that we will consider are the scattering of the initial electron and positron into the final states with one fermion and one antifermion, $e^{-}e^{+}\to f\bar{f}$.  We will ignore the masses of the initial and final-states particles and therefore consider only the processes with $f=\mu, \tau, q$ where $q=u,d,c,s,b$.  The physical process will be identified as $f_{1}+f_{2}\to f_{3}+f_{4}$.  The $s,t$ and $u$-channel are labeled $(1,2),(1,4)$ and $(1,3)$ respectively. 

To make a relatively model-independent phenomenological study of the TeV-scale stringy kinematics, we adopt the parametrised ``bottom-up" approach for the construction of the tree-level string amplitudes~\cite{cim,fhh,bhhm,bfh}.  In this approach, the gauge structure and the assignment of Chan-Paton matrices to the particles are not explicitly specified.  The main requirement is that the relevant string amplitudes reproduce the SM amplitudes in the low energy limit~($s/M^{2}_{S}\to 0$, field theory limit).  This implies that we identify the zeroth-mode string states as the gauge bosons of the SM.  The expression for the open-string 4-fermion tree-level helicity amplitude for the process $\ell_{\alpha}\bar{\ell}_{\beta} \to f_{\alpha}\bar{f}_{\beta}$ with $\alpha, \beta = L, R; \alpha \neq \beta$ is~\cite{bfh}
\begin{eqnarray}
 A_{string}(\ell_{\alpha}\bar{\ell}_{\beta} \rightarrow f_{\alpha}\bar{f}_{\beta})
&=& ig^2_L S(s,t)\frac{t}{s}F_{\alpha\alpha} + ig_L^2T\frac{t}{us}f(s,t,u),
\label{eq:3} \\
 f(s,t,u) &=& uS(s,t) + sS(t,u) + tS(u,s)
\label{eq:1.1}
\end{eqnarray}
where 
\begin{eqnarray}
S(s,t) & = & \frac{\Gamma(1-\alpha's)\Gamma(1-\alpha't)}{\Gamma(1-\alpha's-\alpha't)}
\label{eq:1.2}
\end{eqnarray}
and the Regge slope $\alpha'= M_S^{-2}$.  The interaction factors from the exchange of photon and $Z$~(chiral) are given by 
\begin{equation}
F_{\alpha\beta}=2Q_{\ell}Q_{f}\xw+\frac{s}{s-m^2_{Z}}\ \frac{2g^{\ell}_{\alpha}g^f_{\beta}}{1-\xw}.
\end{equation}
Here 
$\xw=\sin^2\theta_W$ and the $SU(2)_L$ coupling $g_{L}=e/\sin{\theta_{W}}$. The neutral
current couplings are $g^f_{L}=T_{3f}-Q_{f}\xw,\  g^f_{R}=-Q_{f}\xw$.  The Chan-Paton parameter $T$ represents the tree-level stringy interaction which cannot be determined in the field theory limit $s/M^{2}_{S} \to 0$ since $f(s,t,u)\to 0$.  
 The amplitude for other helicity combination $\ell_{\alpha}\bar{\ell}_{\beta} \rightarrow f_{\beta}\bar{f}_{\alpha}$ is given by $t\leftrightarrow u$ and an index exchange in the $F$ factor
\begin{eqnarray}
 A_{string}(\ell_{\alpha}\bar{\ell}_{\beta} \rightarrow f_{\beta}\bar{f}_{\alpha})
&=& ig^2_L S(u,s)\frac{u}{s}F_{\alpha\beta} + ig_L^2T\frac{u}{ts}f(s,t,u).
\label{eq:1.3} 
\end{eqnarray}

Since the Veneziano-like function $S(s,t)$ and $S(u,s)$ are appearing with the chiral couplings~(the factor $F$) in the amplitudes, the piece of the stringy states induced by this function will have the similar chiral couplings to those of the SM.  On the other hand, the purely stringy interaction piece, proportional to $T$, are assumed to be non-chiral and the values of $T$ are set to the same value for all helicity combinations.
            
\section{Low-energy stringy corrections}

For $E\equiv E_{cm}<M_{S}$, we can use Taylor expansion to approximate the leading order stringy corrections to the amplitudes.  They are in the form of the dimension-8 operator
\begin{eqnarray}
S(s,t)& \simeq & 1 - \frac{\pi^2}{6}(\frac{st}{M^{4}_{S}}) \\
f(s,t,u)& \simeq & -\frac{\pi^2}{2}(\frac{stu}{M^{4}_{S}}),
\end{eqnarray}
the amplitude in Eq.~(\ref{eq:3}) becomes
\begin{eqnarray}
A(\ell_{\alpha}\bar{\ell}_{\beta}\to f_{\alpha}\bar{f}_{\beta})& \simeq & ig^2_{L}\left(\frac{t}{s}F_{\alpha\alpha} - \frac{\pi^2}{6}(F_{\alpha\alpha}+3T)\frac{t^2}{M^{4}_{S}}\right). 
\label{eq:12}
\end{eqnarray}
Again, for $\ell_{\alpha}\bar{\ell}_{\beta}\to f_{\beta}\bar{f}_{\alpha}$, the amplitude is $t\leftrightarrow u$ and $F_{\alpha\alpha}\to F_{\alpha\beta}$ of the above.  

The stringy correction can be decomposed into the contribution from the angular momentum states, $j=1,2$.  Using the Wigner functions $d^1_{1,\pm 1}(\cos\theta =1+2t/s)=(1\pm \cos\theta)/2$ and $d^2_{1,\pm 1}(\cos\theta)=\mp (1\pm \cos\theta)(1\mp 2\cos\theta)/2$, 
\begin{eqnarray}
\frac{t^2}{s^2}& = & \frac{3}{4}d^1_{1,-1}-\frac{1}{4}d^2_{1,-1}, \\
\frac{u^2}{s^2}& = & \frac{3}{4}d^1_{1,1}+\frac{1}{4}d^2_{1,1} 
\label{eq:13}
\end{eqnarray}
where $\theta$ is the angle between the incoming electron and the outgoing antifermion in the c.m. frame.
From Eq.~(\ref{eq:12}-\ref{eq:13}), we can see that the couplings of the $j=1,2$ states, proportional to $(F+3T)$, inherit the chirality from the coupling $F$ of the zeroth mode gauge boson exhange in the SM.  This is the distinctive feature of the couplings of the string states in this ``bottom-up" approach.  Because of the chirality of the coupling, the $j=2$ interaction in these stringy amplitudes cannot be described by the effective interaction of the form $-i\lambda T^{\mu\nu}T_{\mu\nu}/M^{4}_{H}$.  As long as we couple the spin-2 state $h^{\mu \nu}$ to the energy-momentum tensor $T_{\mu \nu}$ which does NOT contain information of the chirality of the fermions, the interaction will always be non-chiral.  Therefore, the effective interaction of this kind can never describe the {\it chiral} stringy interaction induced by the worldsheet stringy spin excitations in the string models under consideration~(i.e. the models which address chiral weak interaction).  As a comparison, we give the SM-KK amplitudes where the KK part is induced by the effective interaction of the form $-i\lambda T^{\mu\nu}T_{\mu\nu}/M^{4}_{H}$ as
\begin{eqnarray}
A_{KK}(\ell_{\alpha}\bar{\ell}_{\beta}\to f_{\alpha}\bar{f}_{\beta})& = & ig^2_{L}\frac{t}{s}F_{\alpha\alpha}-2i\lambda \frac{s^2}{M^{4}_{H}}\frac{t}{s}(3+4\frac{t}{s}) \\
A_{KK}(\ell_{\alpha}\bar{\ell}_{\beta}\to f_{\beta}\bar{f}_{\alpha})& = & ig^2_{L}\frac{u}{s}F_{\alpha\beta}+2i\lambda \frac{s^2}{M^{4}_{H}}\frac{u}{s}(3+4\frac{u}{s})
\end{eqnarray}
for $\alpha,\beta =L,R$ and $\alpha \neq \beta$.  The KK-gravitons contributions can be represented by the Wigner functions as 
\begin{eqnarray}
\frac{t}{s}(3+4\frac{t}{s})&=&-d^{2}_{1,-1}(\cos\theta)  \\
\frac{u}{s}(3+4\frac{u}{s})&=& d^{2}_{1,1}(\cos\theta).
\end{eqnarray}
The chiral spin-1 and spin-2 stringy interaction will lead to remarkable and unique phenomenological signatures in the 4-fermion scattering at the electron-positron collider even when compared with the contributions from KK gravitons as we will see in the following.     

\section{The asymmetries}

The left-right~($A_{LR}$) and forward-backward~($A_{FB}$) asymmetries quantify the degrees of the {\it chirality} of the interaction under consideration regardless of the spin of the intermediate states.  On the other hand, the center-edge asymmetry~($A_{CE}$) does NOT contain any information on the chirality of the spin-1 interaction~(at least in the massless limit)~\cite{opp}.  For spin-2, $A_{CE}$ shows dependence on the chirality of the couplings of the intermediate states in the scattering as we will present in Section V.  Therefore, we can use $A_{CE}$ to distinguish NP models with spin-2 mediator from the models mediated by the spin-1 state.  Among the class of models with spin-2 interactions, we can use the $A_{LR}, A_{FB}$ to distinguish one model from another as demonstrated in Fig.~\ref{.1-fig} and Fig.~\ref{10-fig}-\ref{12-fig} for the case of SR versus KK-gravitons.  

\subsection{Left-Right Asymmetries}

\begin{figure}[tb]
\centering
\epsfxsize=5.5in
\hspace*{0in}
\epsffile{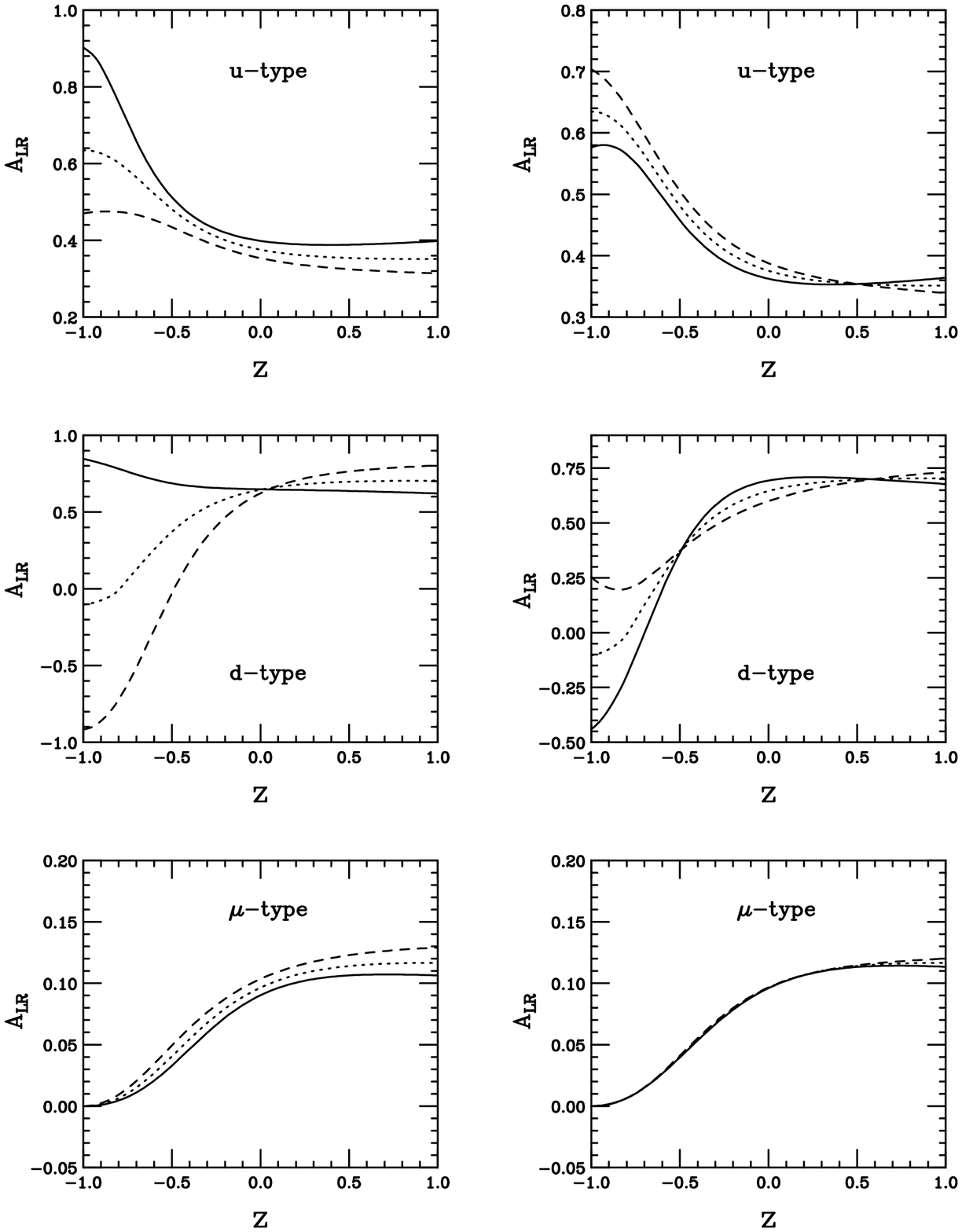}
\caption{
The angular left-right asymmetry for $e^{+}e^{-}\to f\bar{f}$ at $E=500$ GeV.  The dot curve represents the SM. On the left side, the solid(dash) curve represents the SR model with $M_{S}=1.5$ TeV, $T=1(-1)$.  On the right side, the solid(dash) curve represents the KK-graviton model with $M_{H}=2$ TeV, $\lambda=-1(1)$. }
\label{.1-fig}
\end{figure}

The angular left-right asymmetry is defined as a function of $z\equiv -\cos{\theta}$ as 
\begin{eqnarray}
A_{LR}(z) & = & \left( \frac{d\sigma_{L}}{dz}-\frac{d\sigma_{R}}{dz}\right)\Large{/} \left( \frac{d\sigma}{dz}\right) 
\label{eq:2.1} 
\end{eqnarray}
where $\sigma_{L(R)}$ is the cross section of the scattering of the 100\% left(right)-handedly polarized electron beam with the 100\% right(left)-handedly polarized positron beam.  The angular left-right asymmetries induced by the stringy corrections are plotted as in Fig.~\ref{.1-fig} in comparison to the KK-graviton model.  It is interesting to comment that the angular left-right asymmetries induced by the stringy corrections differ significantly from the SM distributions only in the quark~($u$ and $d$) final states.  The difference in $\mu$ is hardly visible.  This feature is similar to the asymmetries induced by the KK gravitons~\cite{hew}.        

\subsection{Forward-Backward Asymmetries}

The forward-backward asymmetry is defined as
\begin{eqnarray}
A_{FB} & = & \frac{N_{F}-N_{B}}{N_{F}+N_{B}} 
\label{eq:2.5}
\end{eqnarray}
where $N_{F(B)}$ is the number of events in the forward(backward) direction.  The numerical values of the unpolarised forward-backward asymmetries for the SM and the stringy amplitudes with $E=500$ GeV, $M_{S}=1.5$ TeV, and $T=1,4$ are
\begin{equation}
A^{string(SM)}_{FB} = \left\{
\begin{array}{l}
0.46, 0.39~(0.49) \qquad ~\mbox{  for $\ell=\mu,\tau$},  \\
0.63, 0.51~(0.62) \qquad ~\mbox{  for $u$},  \\
0.67, 0.60~(0.64) \qquad ~\mbox{  for $d$}.  
\end{array}
\right.
\end{equation}
The deviations of the values with the stringy corrections from the SM values are linearly dependent on $T$ and $(E/M_{S})^4$ at the leading order as we can see from the expression in the polarized beams section with $P_{1(2)}=0$.  This is true only for the scattering at low energies comparing to the string scale.  At higher energies $E\simeq M_{S}$ or around the SRs at $E\simeq \sqrt{n}M_{S}$, the forward-backward asymmetries become very small due to the non-chiral choice of the Chan-Paton parameters $T$ which efficiently dilutes the chirality of the interaction~\cite{bfh}.     

\subsection{Center-Edge Asymmetries}

The center-edge asymmetry is defined as a function of the cut of the central region $z^{*}$ as

\begin{eqnarray}
A_{CE} & = &\frac{1}{\sigma}\left[ \left( \int^{z^{*}}_{-z^{*}}-(\int^{-z^{*}}_{-1}+\int^{1}_{z^{*}}) \right) \frac{d\sigma}{dz} dz  \right]. 
\label{eq:2.10}
\end{eqnarray}
The deviations of the center-edge asymmetry from the SM values of the unpolarised beams, $\Delta A^{string}_{CE}=A^{string}_{CE}-A^{SM}_{CE}$, are plotted with respect to $z^{*}$ in Fig.~\ref{4-fig}.  A few interesting remarks are worthwhile making.  There is distinctive feature between $A_{CE}$ of the lepton $\mu$ and the quarks, $u$ and $d$.  The effect of the non-chiral purely stringy interaction represented by the value of $T$ is opposite for $\mu$ and $u$.  On the other hand, the features of $\mu$ and $d$-type quark are, for the small value of $T(\lesssim 1)$, roughly the same.  For large value of $T(\simeq 4)$, the deviation from the SM of the $d$-type becomes negative and appears very distinctive from the corresponding value of the $\mu$.  This strange behaviour is originated from the competing contributions between $T^2$ appearing in the $O((E/M_{S})^8)$ terms and the terms of order $O((E/M_{S})^4)$.       
 
It should be emphasized that the numerical results as in Fig.~\ref{4-fig} are calculated using full expression upto the order of $O((E/M_{S})^8)$.  They are different from the results obtained from the approximation using the leading order upto $O((E/M_{S})^4)$ as is given in Section V.  The difference becomes very obvious in the $d$-type quark with high value of $T(\simeq 4)$.   

\section{Partially Polarized beams}

Let $P_{1(2)}$ be the degrees of the longitudinal polarization of the $e^{-(+)}$ beam defined as~($i=1,2$)
\begin{eqnarray}
P_{i} & = & \frac{N^{i}_{R}-N^{i}_{L}}{N^{i}_{R}+N^{i}_{L}} 
\end{eqnarray}
where $N^{i}_{L(R)}$ is the number of particle $i$ with the left(right)-handed helicity in the beams.  The polarized differential cross section can be expressed as
\begin{eqnarray}
\frac{d\sigma^{pol}}{dz} & = & \left[ D_{L}\left( \frac{d\sigma_{LL}}{dz}+\frac{d\sigma_{LR}}{dz} \right)+D_{R}\left( \frac{d\sigma_{RR}}{dz}+\frac{d\sigma_{RL}}{dz} \right) \right] 
\label{eq:3.1}
\end{eqnarray}
where

\begin{equation}
D_{L,R} = \frac{1}{4}(1\mp P_{1})(1\pm P_{2}).
\end{equation}
Upper(lower) signs are for $L(R)$.  $D_{L(R)}$ represents the scattering involving the left(right)-handed electron.    

\subsection{The partially polarized left-right asymmetries}

In practice, the observable left-right asymmetry is defined with respect to the partially polarized beams of electron and positron by taking difference of the total number of events when the polarizations of the beams are inverted.  It is therefore given by 
\begin{eqnarray}
A_{obs}& = & -P_{comb}A_{LR}
\end{eqnarray}
where $P_{comb}=(D_{R}-D_{L})/(D_{R}+D_{L})$.  For $(P_{1},P_{2})=(-0.8,0)$, $P_{comb}$ is $-0.8$, while $P_{comb}$ becomes $-0.946$ when $(P_{1},P_{2})=(-0.8,0.6)$.  The full expressions of $A_{LR}(z)$ for the KK-gravitons and SR models are given in the Appendix.

\subsection{The polarized forward-backward asymmetries}

With the partially polarised beams of electron and positron, we can calculate $A_{FB}$ using Eq.~(\ref{eq:3.1}).  The full expressions for the asymmetries induced by KK gravitons and SR are given in the Appendix.  Up to the leading order of $O((E/M_{S})^4)$ in the energy Taylor expansion, the polarized forward-backward asymmetry induced by the stringy corrections is 
\begin{eqnarray}
A^{pol,string}_{FB}& = & A^{pol,SM}_{FB}+\Delta A^{pol,string}_{FB}
\end{eqnarray}

\begin{eqnarray*}
\Delta A^{pol,string}_{FB} & = & -\frac{\pi^{2}}{6}\left(\frac{E}{M_{S}}\right)^{4} \\
& \times & \left[ \left(\frac{1}{4} - 3T \frac{D_{\alpha}\Sigma F_{\alpha}}{d_{SM}}\right) A^{pol,SM}_{FB} + \left(\frac{21T}{8}\right)\frac{D_{L}(F_{LL}-F_{LR})+D_{R}(F_{RR}-F_{RL})}{d_{SM}} \right]
\label{eq:3.3}  
\end{eqnarray*}
where the Einstein summation convention is implied for $\alpha =L,R$.  $\Sigma F_{L}\equiv F_{LL}+F_{LR}$ and $\Sigma F^{2}_{L}\equiv F^{2}_{LL}+F^{2}_{LR}$.  The other combination is obtained by exchanging $L\leftrightarrow R$.  The SM value is given by
\begin{equation}
A^{pol,SM}_{FB} = \frac{n_{SM}}{d_{SM}}
\end{equation}  
with $n_{SM}=(D_{L}(F^{2}_{LL}-F^{2}_{LR})+D_{R}(F^{2}_{RR}-F^{2}_{RL}))/2$ and $d_{SM}=2(D_{\alpha}\Sigma F^{2}_{\alpha})/3$.

\begin{figure}
\centering
\epsfxsize=4.0in
\hspace*{0in}
\epsffile{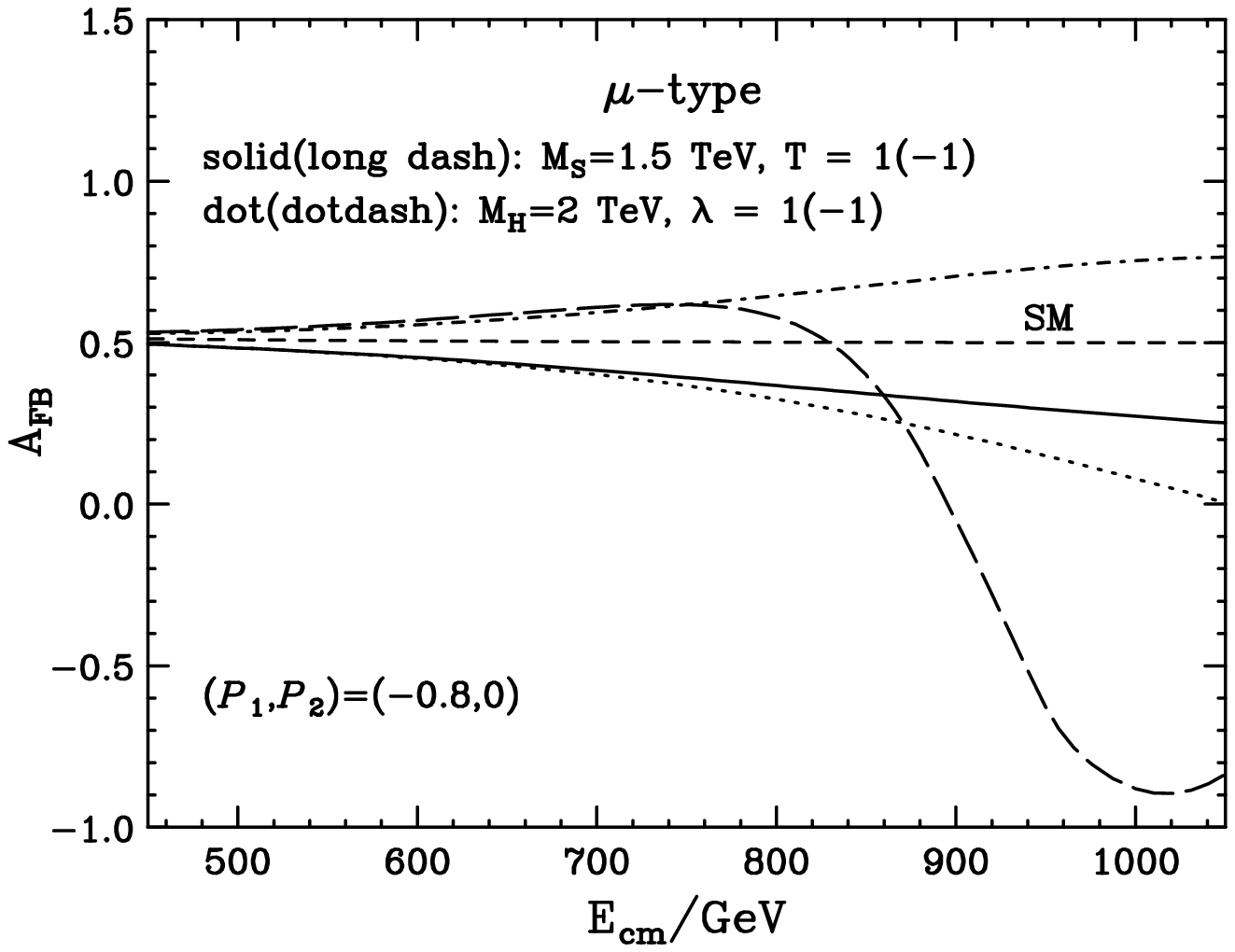}
\caption{The forward-backward asymmetry $A_{FB}$ for $f=\mu,\tau$ induced by stringy corrections and KK-gravitons interfering with the tree-level SM contribution when the electron and positron beams have polarization $(P_{1},P_{2})=(-0.8,0)$.  The SM value is also given for comparison as an almost constant dash line.}
\label{10-fig}
\end{figure}

\begin{figure}
\centering
\epsfxsize=4.0in
\hspace*{0in}
\epsffile{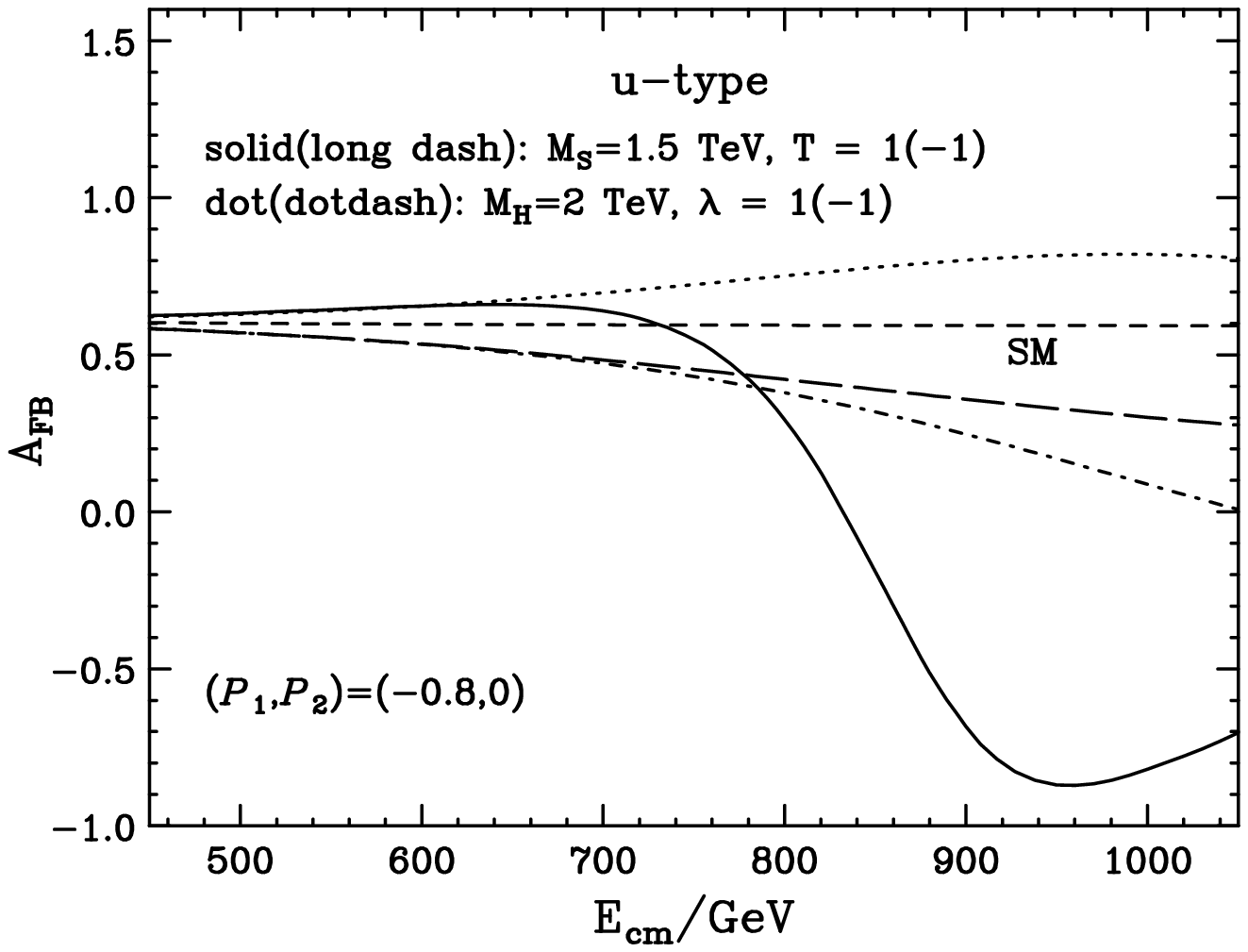}
\caption{The forward-backward asymmetry $A_{FB}$ for $f=u,c$ induced by stringy corrections and KK-gravitons interfering with the tree-level SM contribution when the electron and positron beams have polarization $(P_{1},P_{2})=(-0.8,0)$.  The SM value is also given for comparison as an almost constant dash line.}
\label{11-fig}
\end{figure}

\begin{figure}
\centering
\epsfxsize=4.0in
\hspace*{0in}
\epsffile{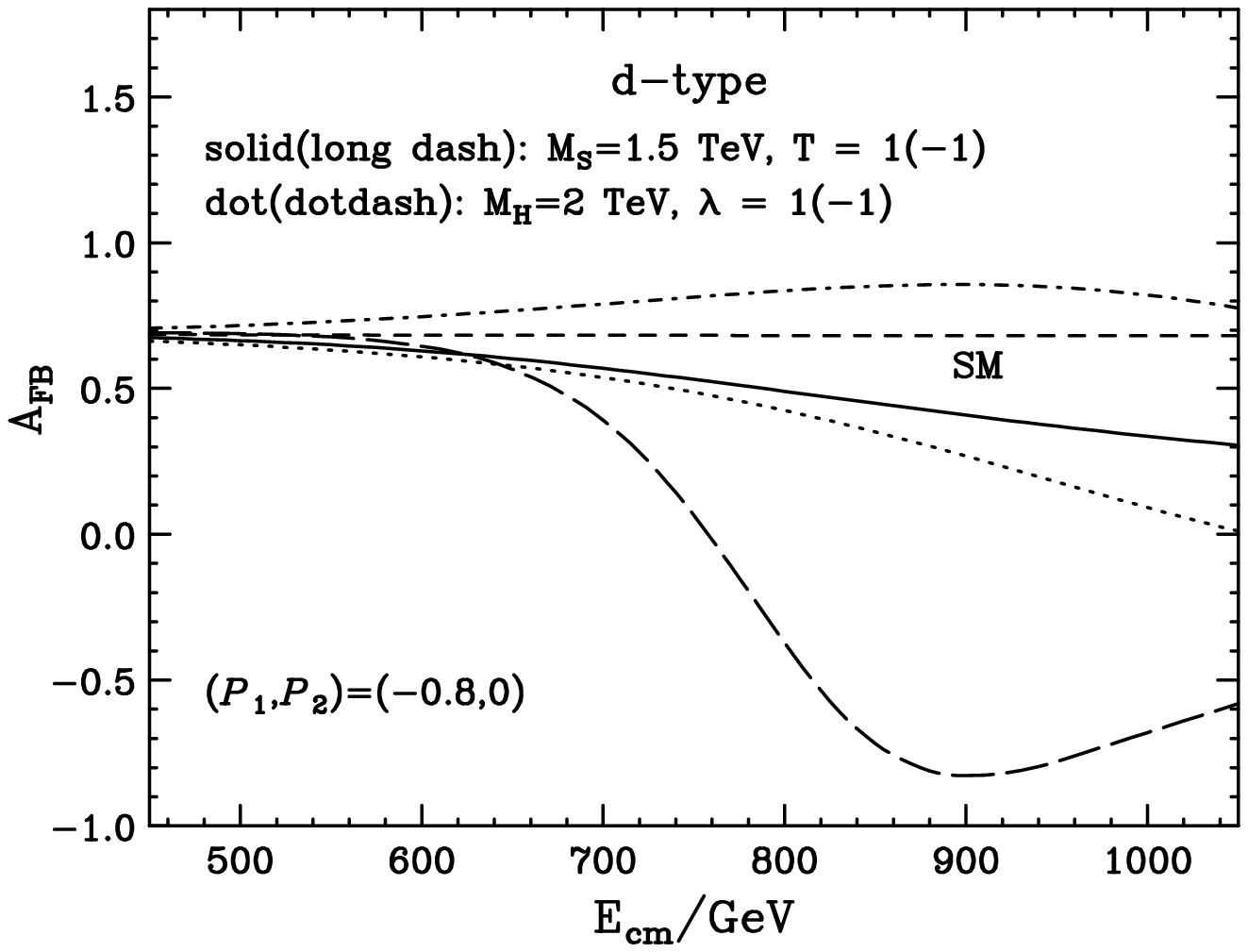}
\caption{The forward-backward asymmetry $A_{FB}$ for $f=d,s,b$ induced by stringy corrections and KK-gravitons interfering with the tree-level SM contribution when the electron and positron beams have polarization $(P_{1},P_{2})=(-0.8,0)$.  The SM value is also given for comparison as an almost constant dash line.}
\label{12-fig}
\end{figure}

Fig.~\ref{10-fig}-\ref{12-fig} show $A_{FB}$ induced by KK gravitons and SR in comparison to the SM values with respect to the CM energy.  Assuming the string scale $M_{S}=1.5$ TeV and the effective quantum gravity scale $M_{H}=2$ TeV, the differences between the asymmetries induced by the two models appear at higher CM energies even in the case when they are indistinguishable at the low CM energies.  This is due to the fact that while the {\it chirality} of the stringy interaction keeps the terms of both order $O((E/M_{S})^4)$ and $O((E/M_{S})^8)$ in the expression of $A_{FB}$, the non-chiral graviton interaction, on the other hand, keeps only the term of order $O((E/M_{H})^4)$ in the numerator $n_{FB}$, and only term of order $O((E/M_{H})^8)$ in the denominator $d_{FB}$~(see Appendix).  This leads to distinguishable aspects of the two models.
   

\subsection{The polarized center-edge asymmetries}

\begin{figure}
\centering
\epsfxsize=4.0in
\hspace*{0in}
\epsffile{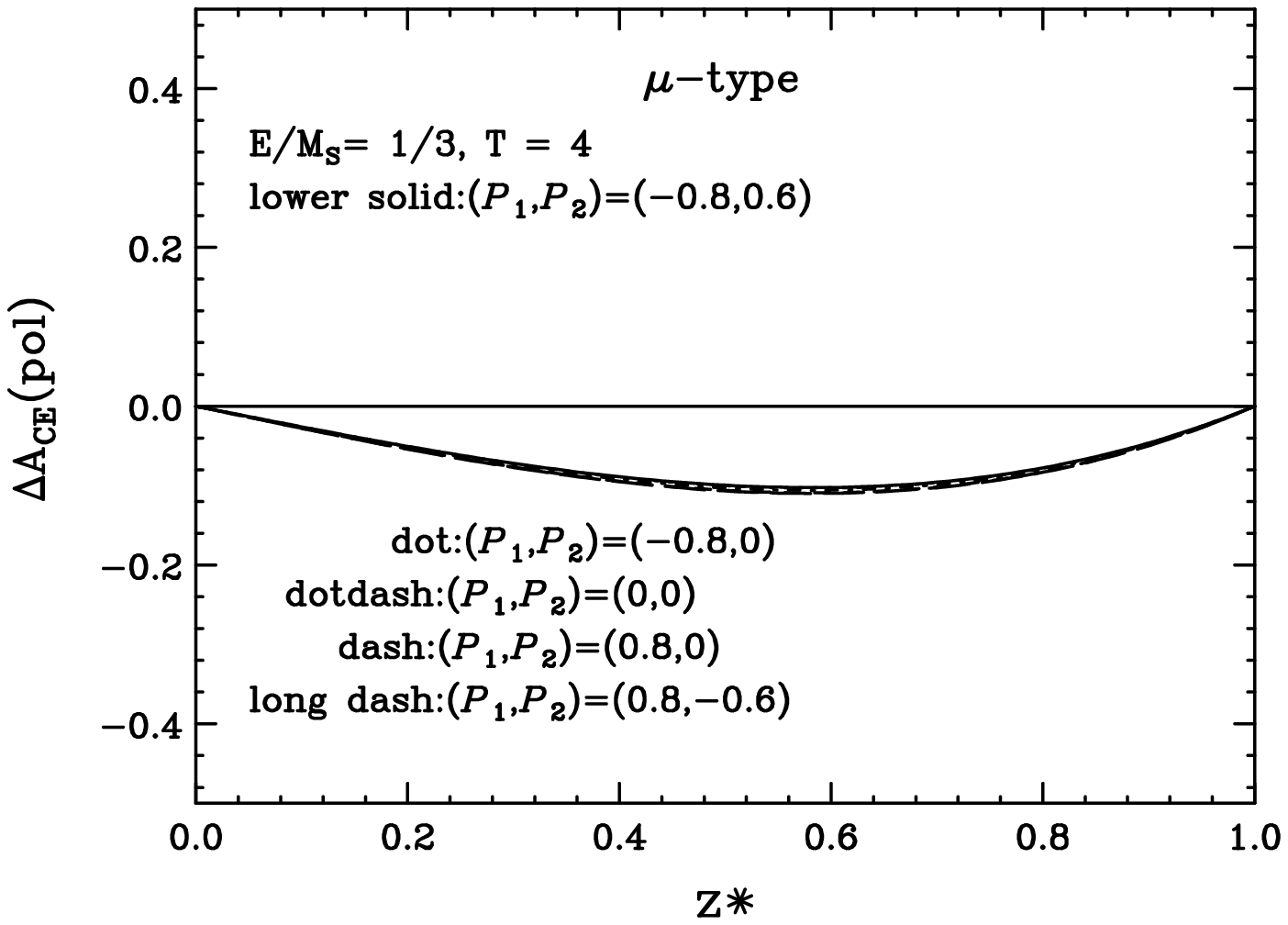}
\caption{The deviations from the SM value of the center-edge asymmetry induced by stringy corrections $\Delta A_{CE}=A_{CE}-A_{SM}$ for the $f=\mu,\tau$ when $E/M_{S}=3, T=4$ for various degrees of polarization of the electron~($P_{1}$) and positron~($P_{2}$) beams.}
\label{7-fig}
\end{figure}

\begin{figure}
\centering
\epsfxsize=4.0in
\hspace*{0in}
\epsffile{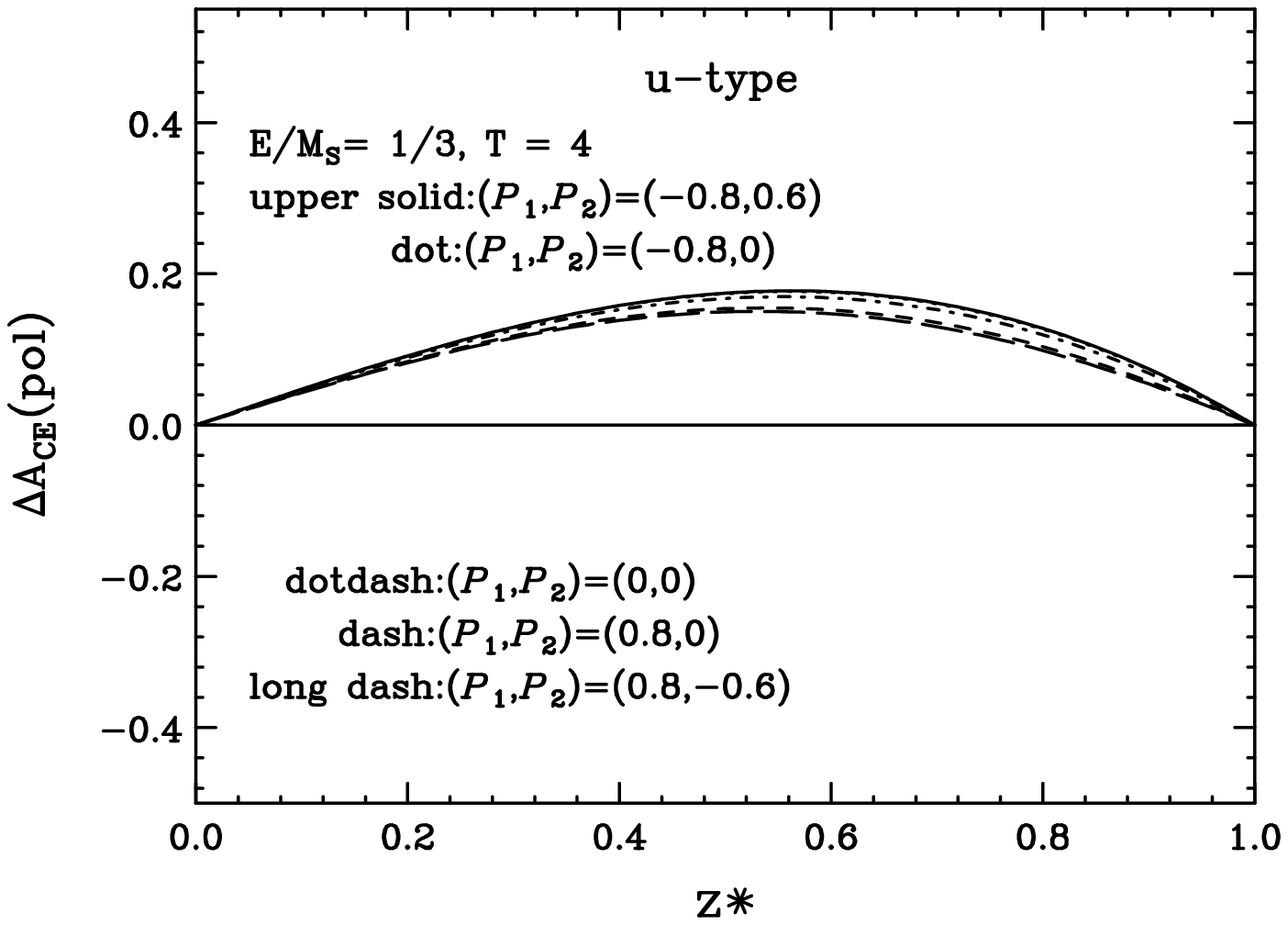}
\caption{The deviations from the SM value of the center-edge asymmetry induced by stringy corrections $\Delta A_{CE}=A_{CE}-A_{SM}$ for the $f=u,c$ when $E/M_{S}=3, T=4$ for various degrees of polarization of the electron~($P_{1}$) and positron~($P_{2}$) beams.}
\label{8-fig}
\end{figure}

\begin{figure}
\centering
\epsfxsize=4.0in
\hspace*{0in}
\epsffile{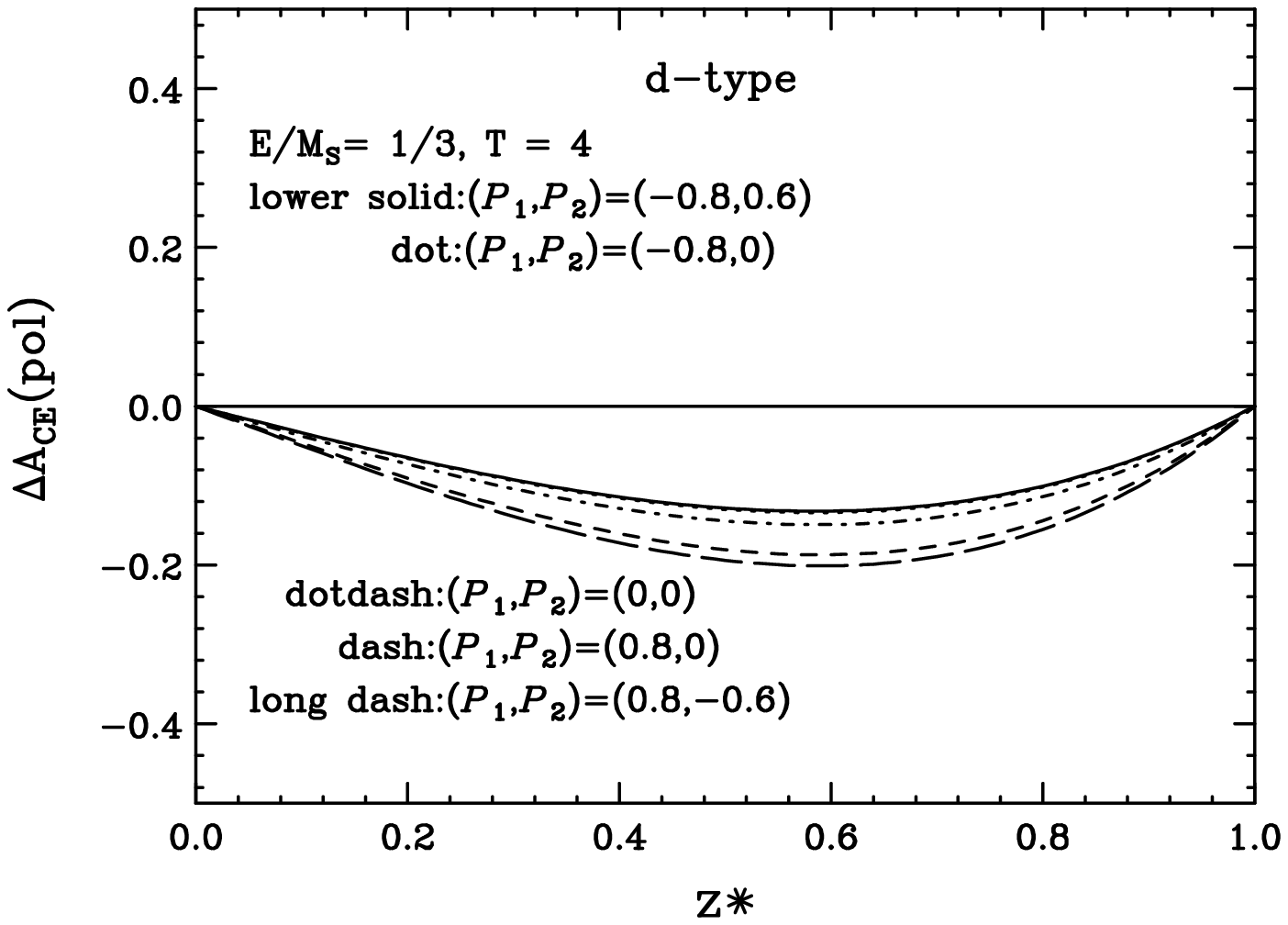}
\caption{The deviations from the SM value of the center-edge asymmetry induced by stringy corrections $\Delta A_{CE}=A_{CE}-A_{SM}$ for the $f=d,s,b$ when $E/M_{S}=3, T=4$ for various degrees of polarization of the electron~($P_{1}$) and positron~($P_{2}$) beams.}
\label{9-fig}
\end{figure}

With the partially polarised beams of electron and positron, we can calculate $A_{CE}$ using Eq.~(\ref{eq:3.1}).  The full expressions for the asymmetries induced by KK gravitons and SR are given in the Appendix.  The deviations from the SM values for various polarizations of the beams for the stringy model are plotted as in Fig.~\ref{7-fig}-\ref{9-fig}.  Up to the leading order of $O((E/M_{S})^4)$ in the energy Taylor expansion, the polarized center-edge asymmetry induced by the stringy corrections is
\begin{eqnarray}
A^{pol,string}_{CE}& = & A^{pol,SM}_{CE}+\Delta A^{pol,string}_{CE}
\end{eqnarray}

\begin{equation}
\Delta A^{pol,string}_{CE} = -(\frac{3}{4})\frac{\pi^{2}}{6}\left( \frac{E}{M_{S}}\right)^{4}\left( 1+3T \frac{D_{\alpha}\Sigma F_{\alpha}}{D_{\alpha}\Sigma F^{2}_{\alpha}} \right) z^{*}(1-z^{*2}).
\label{eq:3.5}  
\end{equation}
where the SM value is given by
\begin{eqnarray}
A^{pol,SM}_{CE}& = & \frac{1}{2}z^{*}(3+z^{*2})-1.
\label{eq:c3}
\end{eqnarray}
Eq.~(\ref{eq:c3}) is unique for the spin-1 contribution in the 4-fermion scattering.  Any NP particles with spin-1 will not change this $z^{*}$-dependence.  Remarkably, the spin-2 contributions, either in the form of KK~\cite{opp} or the string states~(Eq.~(\ref{eq:3.5})), will induce the deviations from this SM value proportional to $z^{*}(1-z^{*2})$. 

\begin{figure}[tb]
\centering
\epsfxsize=5.5in
\hspace*{0in}
\epsffile{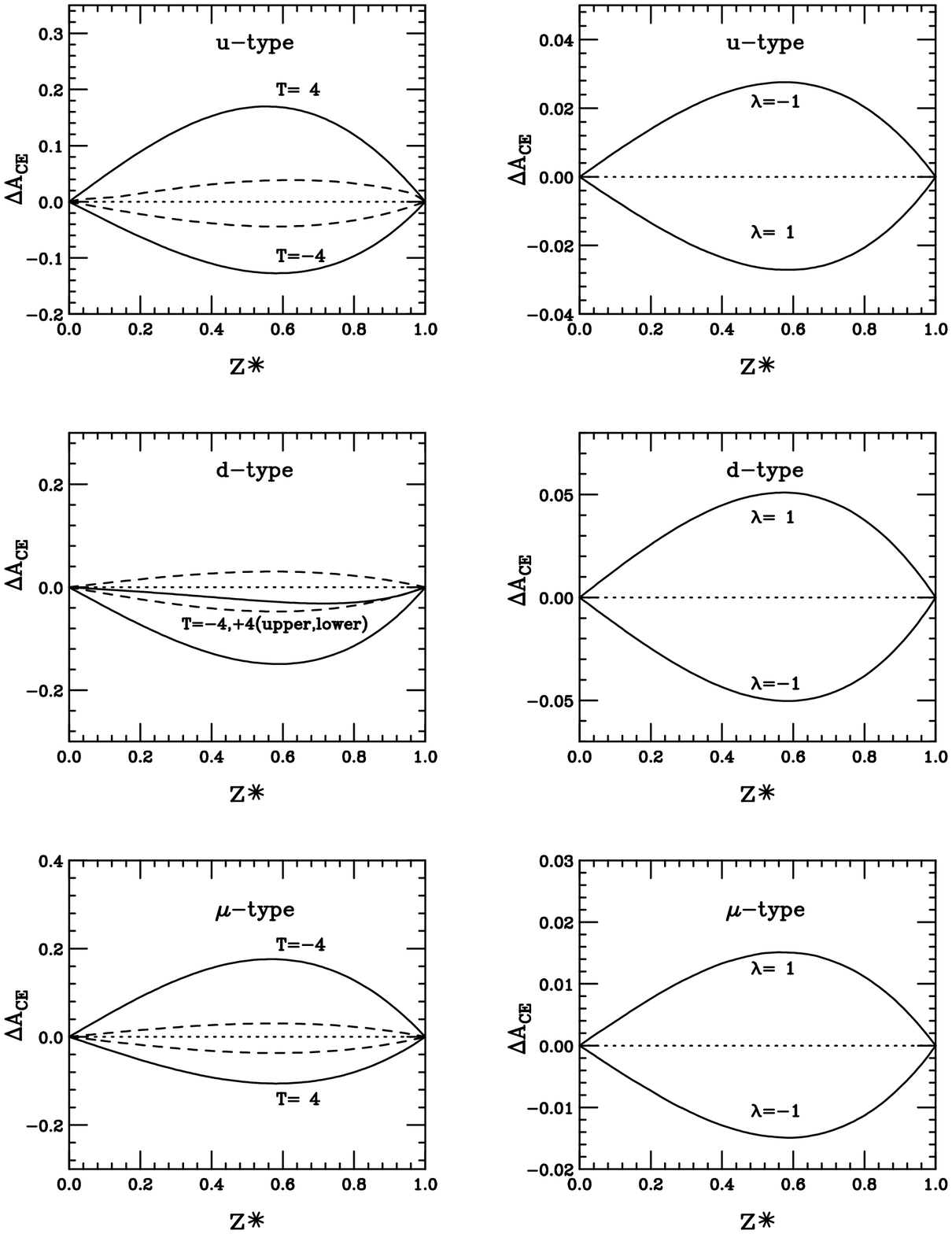}
\caption{The deviations of the center-edge asymmetry from the SM values for the scattering $e^{+}e^{-}\to f\bar{f}$.  The left side is the SR model with $E/M_{S}=1/3$.  Solid(dash) curve represents $|T|=4(1)$.  The right side is the KK-graviton model with $E/M_{H}=1/4$. }
\label{4-fig}
\end{figure}

Similar to the $A_{FB}$ case, the {\it chirality} of the stringy interaction keeps terms of both order $O((E/M_{S})^4)$ and $O((E/M_{S})^8)$ in the expression of $A_{CE}(z^{*})$ while in the case of KK gravitons, there is NO terms of order $O((E/M_{H})^4)$ in the constant term $c_{0}$ and the denominator $d_{CE}$.  The differences become obvious when $|T|$ is relatively large~($\simeq 4$) as shown in Fig.~\ref{4-fig}.

\section{Conclusions}
\label{conclusions-sec}

We have constructed and approximated the tree-level stringy amplitudes of the scattering $e^{+}e^{-}\to f\bar{f}$.  The low-energy stringy corrections for the 4-fermion processes contain both spin-1 and spin-2 contributions with the chiral couplings inherited from the zeroth mode states identified with the gauge bosons in the SM.  The chirality of the couplings is diluted by the non-chiral choice of the purely-stringy piece of the stringy interaction, characterized by $T$.  The contributions from both stringy spin-1 and spin-2 are of dimension-8 in nature.  The low-energy dimension-8 spin-1 contribution is remarkably distinctive from the dimension-6 contributions in other $Z'$ models.  The chirality of stringy spin-2 interaction also leads to unique phenomenological features distinguishable from the non-chiral spin-2 interaction induced by the KK-gravitons~(or massive gravitons).   

Then we investigated the signatures of the TeV-scale string model at the $e^{+}e^{-}$ collider in comparison to the KK-gravitons using angular left-right, forward-backward, and center-edge asymmetries.  The deviations of the asymmetries from the SM values are investigated separately for each model.  All asymmetries show significant differences between the low-energy corrections of the two models.  For the $e^{+}e^{-}$ collider with variable center-of-mass energies from 500 to 1000 GeV and assuming $M_{S(H)}=1.5(2)$ TeV, the forward-backward asymmetries show drastic differences between stringy signals and the KK-graviton ones.  The center-edge asymmetries also show significant differences between the two models if the Chan-Paton parameter $|T|$~(representing purely stringy piece of the interaction) in the SR model is sufficiently large~($\simeq 4$).  

The origin of the differences between the two models is mainly~(another reason is the fact that SR also has spin-1 contribution in addition to spin-2) the {\it chirality} of the interactions.  While the string interaction is chiral, the interaction induced by the KK gravitons, couple to the energy-momentum tensor $T_{\mu \nu}$, is non-chiral.  As we can see in the Appendix from the full espressions of $A_{LR}(z), A_{FB}, A_{CE}(z^{*})$, the chirality of the stringy interaction keeps almost all of the terms of order $O((E/M_{S})^4)$ and $O((E/M_{S})^8)$ while the non-chirality of the interaction of the KK gravitons eliminates certain terms of order $O((E/M_{H})^4)$ and $O((E/M_{H})^8)$ in the asymmetries.  Specifically, the $a_{1}$ term in $A_{LR}(z)$ contains only the SM term in the case of KK gravitons, in contrast to the stringy case.  In $A_{FB}$, as discussed above, $n_{FB}(d_{FB})$ in the case of KK gravitons does not contain the term of order $O((E/M_{H})^8)(O((E/M_{H})^4))$.  For $A_{CE}(z^{*})$, $c_{0}, d_{CE}$ does not contain terms of order of $O((E/M_{H})^4)$ in the case of KK gravitons.  This is in contrast to the processes where the stringy interaction is non-chiral as well as having only the spin-2 contributions such as in the scattering $f\bar{f}, \gamma \gamma\to \gamma \gamma$; in which case the two models give exactly the same low-energy signatures~\cite{pb}.  
   
In the intermediate energy range~($m_{Z}\ll E<M_{S(H)}$) with $s/(s-m^2_{Z})\approx 1.$, since the deviations induced by KK gravitons and SR depend only on $(E/M)$, the results in this article therefore are also valid for the CLIC with center-of-mass energies 3 to 6 TeV with $M_{S(H)}=9(12)$ TeV.        

\section*{Acknowledgments}
\indent
I would like to thank Tao Han for helpful discussions.  This work was supported in part by the U.S. Department of Energy under contract number DE-FG02-01ER41155.

\appendix

\section{Angular Left-Right Asymmetry}

\begin{eqnarray}
A_{LR}(z)& = & \frac{a_{0}+a_{1}z+a_{2}z^{2}+a_{3}z^{3}+a_{4}z^4}{b_{0}+b_{1}z+b_{2}z^{2}+b_{3}z^{3}+b_{4}z^4}
\end{eqnarray}

\subsection{\underline{KK-graviton}}
\begin{eqnarray*}
a_{0}& = & \left( \frac{4\pi \alpha}{x_{W}}\right)^{2}\frac{1}{4}(\Sigma F^{2}_{L}-\Sigma F^{2}_{R}) \\
     & + & \lambda \left( \frac{4\pi \alpha}{x_{W}}\right)\left( \frac{E}{M_{H}}\right)^4((F_{LL}-F_{LR})-(F_{RR}-F_{RL})) \\
a_{1}& = & -\left( \frac{4\pi \alpha}{x_{W}}\right)^{2}\frac{1}{2}(-F_{LL}^{2}+F_{LR}^{2}+F_{RR}^{2}-F_{RL}^{2}) \\
a_{2}& = & \left( \frac{4\pi \alpha}{x_{W}}\right)^{2}\frac{1}{4}(\Sigma F_{L}^{2}-\Sigma F_{R}^{2}) \\
     & - & 3\lambda \left( \frac{4\pi \alpha}{x_{W}}\right)\left( \frac{E}{M_{H}}\right)^4((F_{LL}-F_{LR})-(F_{RR}-F_{RL})) \\
a_{3}& = & -2\lambda\left( \frac{4\pi \alpha}{x_{W}}\right)\left( \frac{E}{M_{H}}\right)^4(\Sigma F_{L}-\Sigma F_{R}) \\
a_{4}& = & 0
\end{eqnarray*}

\begin{eqnarray*}
b_{0}& = & \left( \frac{4\pi \alpha}{x_{W}}\right)^{2}\frac{1}{4}(\Sigma F^{2}_{L}+\Sigma F^{2}_{R}) \\
     & + & \lambda \left( \frac{4\pi \alpha}{x_{W}}\right)\left( \frac{E}{M_{H}}\right)^4((F_{LL}-F_{LR})+(F_{RR}-F_{RL})) \\
     & + & 4\lambda^{2}\left( \frac{E}{M_{H}}\right)^{8}  \\
b_{1}& = & -\left( \frac{4\pi \alpha}{x_{W}}\right)^{2}\frac{1}{2}(-F_{LL}^{2}+F_{LR}^{2}-F_{RR}^{2}+F_{RL}^{2}) \\
b_{2}& = & \left( \frac{4\pi \alpha}{x_{W}}\right)^{2}\frac{1}{4}(\Sigma F^{2}_{L}+\Sigma F^{2}_{R}) \\
     & - & 3\lambda \left( \frac{4\pi \alpha}{x_{W}}\right)\left( \frac{E}{M_{H}}\right)^4((F_{LL}-F_{LR})+(F_{RR}-F_{RL})) \\
     & - & 12\lambda^{2}(\frac{E}{M_{H}})^{8}  \\
b_{3}& = & -2\lambda\left( \frac{4\pi \alpha}{x_{W}}\right)\left( \frac{E}{M_{H}}\right)^4(\Sigma F_{L}+\Sigma F_{R}) \\
b_{4}& = & 16\lambda^{2}\left( \frac{E}{M_{H}}\right)^{8}
\end{eqnarray*}

\subsection{\underline{SR}}
\begin{eqnarray*}
a_{0}& = & \frac{1}{4}(\Sigma F^{2}_{L}-\Sigma F^{2}_{R}) \\
     & + & \frac{\pi^{2}}{24}\left( \frac{E}{M_{S}}\right)^{4}\left( (\Sigma F^{2}_{L}+3T\Sigma F_{L})-(\Sigma F^{2}_{R}+3T\Sigma F_{R}) \right) \\
     & + & \frac{\pi^{4}}{576}\left( \frac{E}{M_{S}}\right)^{8}\left( (\Sigma F^{2}_{L}+6T\Sigma F_{L})-(\Sigma F^{2}_{R}+6T\Sigma F_{R}) \right) \\
a_{1}& = & \frac{1}{2}\left( (F^{2}_{LL}-F^{2}_{LR})-(F^{2}_{RR}-F^{2}_{RL})\right) \\
     & + & \frac{\pi^{2}}{8}\left( \frac{E}{M_{S}}\right)^{4}\left( (F^{2}_{LL}-F^{2}_{LR}+3T(F_{LL}-F_{LR}))-(F^{2}_{RR}-F^{2}_{RL}+3T(F_{RR}-F_{RL})) \right) \\
     & + & \frac{\pi^{4}}{144}\left( \frac{E}{M_{S}}\right)^{8}\left( (F^{2}_{LL}-F^{2}_{LR}+6T(F_{LL}-F_{LR}))-(F^{2}_{RR}-F^{2}_{RL}+6T(F_{RR}-F_{RL})) \right) \\
a_{2}& = & \frac{1}{4}(\Sigma F^{2}_{L}-\Sigma F^{2}_{R}) \\
     & + & \frac{\pi^{2}}{8}\left( \frac{E}{M_{S}}\right)^{4}\left( (\Sigma F^{2}_{L}+3T\Sigma F_{L})-(\Sigma F^{2}_{R}+3T\Sigma F_{R}) \right) \\
     & + & \frac{\pi^{4}}{96}\left( \frac{E}{M_{S}}\right)^{8}\left( (\Sigma F^{2}_{L}+6T\Sigma F_{L})-(\Sigma F^{2}_{R}+6T\Sigma F_{R}) \right) \\
a_{3}& = & \frac{\pi^{2}}{3}\left( \frac{E}{M_{S}}\right)^{4}\left( (F^{2}_{LL}-F^{2}_{LR}+3T(F_{LL}-F_{LR}))-(F^{2}_{RR}-F^{2}_{RL}+3T(F_{RR}-F_{RL})) \right) \\
     & + & \frac{\pi^{4}}{144}\left( \frac{E}{M_{S}}\right)^{8}\left( (F^{2}_{LL}-F^{2}_{LR}+6T(F_{LL}-F_{LR}))-(F^{2}_{RR}-F^{2}_{RL}+6T(F_{RR}-F_{RL})) \right) \\
a_{4}& = & \frac{\pi^{4}}{576}\left( \frac{E}{M_{S}}\right)^{8}\left( (\Sigma F^{2}_{L}+6T\Sigma F_{L})-(\Sigma F^{2}_{R}+6T\Sigma F_{R}) \right) \\
\end{eqnarray*}

\begin{eqnarray*}
b_{0}& = &  \frac{1}{4}(\Sigma F^{2}_{L}+\Sigma F^{2}_{R}) \\
     & + & \frac{\pi^{2}}{24}\left( \frac{E}{M_{S}}\right)^{4}\left( (\Sigma F^{2}_{L}+3T\Sigma F_{L})+(\Sigma F^{2}_{R}+3T\Sigma F_{R}) \right) \\
     & + & \frac{\pi^{4}}{576}\left( \frac{E}{M_{S}}\right)^{8}\left( (\Sigma F^{2}_{L}+6T\Sigma F_{L})+(\Sigma F^{2}_{R}+6T\Sigma F_{R})+36T^{2} \right) \\
b_{1}& = & \frac{1}{2}\left( (F^{2}_{LL}-F^{2}_{LR})+(F^{2}_{RR}-F^{2}_{RL}) \right) \\
     & + & \frac{\pi^{2}}{8}\left( \frac{E}{M_{S}}\right)^{4}\left( (F^{2}_{LL}-F^{2}_{LR}+3T(F_{LL}-F_{LR}))+(F^{2}_{RR}-F^{2}_{RL}+3T(F_{RR}-F_{RL})) \right) \\
     & + & \frac{\pi^{4}}{144}\left( \frac{E}{M_{S}}\right)^{8}\left( (F^{2}_{LL}-F^{2}_{LR}+6T(F_{LL}-F_{LR}))+(F^{2}_{RR}-F^{2}_{RL}+6T(F_{RR}-F_{RL})) \right) \\
b_{2}& = & \frac{1}{4}(\Sigma F^{2}_{L}+\Sigma F^{2}_{R}) \\
     & + & \frac{\pi^{2}}{8}\left( \frac{E}{M_{S}}\right)^{4}\left( (\Sigma F^{2}_{L}+3T\Sigma F_{L})+(\Sigma F^{2}_{R}+3T\Sigma F_{R}) \right) \\
     & + & \frac{\pi^{4}}{96}\left( \frac{E}{M_{S}}\right)^{8}\left( (\Sigma F^{2}_{L}+6T\Sigma F_{L})+(\Sigma F^{2}_{R}+6T\Sigma F_{R})+36T^{2} \right) \\
b_{3}& = & \frac{\pi^{2}}{3}\left( \frac{E}{M_{S}}\right)^{4}\left( (F^{2}_{LL}-F^{2}_{LR}+3T(F_{LL}-F_{LR}))+(F^{2}_{RR}-F^{2}_{RL}+3T(F_{RR}-F_{RL})) \right) \\
     & + & \frac{\pi^{4}}{144}\left( \frac{E}{M_{S}}\right)^{8}\left( (F^{2}_{LL}-F^{2}_{LR}+6T(F_{LL}-F_{LR}))+(F^{2}_{RR}-F^{2}_{RL}+6T(F_{RR}-F_{RL})) \right) \\
b_{4}& = & \frac{\pi^{4}}{576}\left( \frac{E}{M_{S}}\right)^{8}\left( (\Sigma F^{2}_{L}+6T\Sigma F_{L})+(\Sigma F^{2}_{R}+6T\Sigma F_{R})+36T^{2} \right) \\
\end{eqnarray*}

\section{Forward-Backward Asymmetry}

\begin{eqnarray}
A_{FB}& = & \frac{n_{FB}}{d_{FB}}
\end{eqnarray}
\subsection{\underline{KK-graviton}}
\begin{eqnarray*}
n_{FB}& = & \frac{1}{2}\left( D_{L}(F^{2}_{LL}-F^{2}_{LR})+D_{R}(F^{2}_{RR}-F^{2}_{RL}) \right)-\frac{\lambda x_{W}}{4\pi \alpha}\left( \frac{E}{M_{H}}\right)^{4}(D_{L}\Sigma F_{L}+D_{R}\Sigma F_{R}) \\
d_{FB}& = & \frac{2}{3}(D_{L}\Sigma F^{2}_{L}+D_{R}\Sigma F^{2}_{R})+\frac{16}{5}\frac{\lambda^{2}x^{2}_{W}}{(4\pi \alpha)^2}(D_{L}+D_{R})\left( \frac{E}{M_{H}}\right)^8
\end{eqnarray*}
\subsection{\underline{SR}}
\begin{eqnarray*}
n_{FB}& = &  \frac{1}{2}\left( D_{L}(F^{2}_{LL}-F^{2}_{LR})+D_{R}(F^{2}_{RR}-F^{2}_{RL}) \right) \\
     & + & \frac{7\pi^{2}}{48}\left( \frac{E}{M_{S}}\right)^{4}\left( D_{L}(F^{2}_{LL}-F^{2}_{LR}+3T(F_{LL}-F_{LR}))+D_{R}(F^{2}_{RR}-F^{2}_{RL}+3T(F_{RR}-F_{RL})) \right) \\
     & + & \frac{\pi^{4}}{96}\left( \frac{E}{M_{S}}\right)^{8}\left( D_{L}(F^{2}_{LL}-F^{2}_{LR}+6T(F_{LL}-F_{LR}))+D_{R}(F^{2}_{RR}-F^{2}_{RL}+6T(F_{RR}-F_{RL})) \right) \\
d_{FB}& = & \frac{2}{3}(D_{L}\Sigma F^{2}_{L}+D_{R}\Sigma F^{2}_{R}) \\
     & + & \frac{\pi^{2}}{6}\left( \frac{E}{M_{S}}\right)^{4}\left( D_{L}(\Sigma F^{2}_{L}+3T\Sigma F_{L})+D_{R}(\Sigma F^{2}_{R}+3T\Sigma F_{R}) \right) \\
     & + & \frac{\pi^{4}}{90}\left( \frac{E}{M_{S}}\right)^{8}\left( D_{L}(\Sigma F^{2}_{L}+6T\Sigma F_{L})+D_{R}(\Sigma F^{2}_{R}+6T\Sigma F_{R})+18T^{2}(D_{L}+D_{R}) \right) 
\end{eqnarray*}
\section{Center-Edge Asymmetry}

\begin{eqnarray}
A_{CE}(z^{*})& = & \frac{c_{0}+c_{1}z^{*}+c_{2}z^{*2}+c_{3}z^{*3}+c_{4}z^{*4}+c_{5}z^{*5}}{d_{CE}}
\end{eqnarray}
\subsection{\underline{KK-graviton}}
\begin{eqnarray*}
c_{0} & = & -\frac{2}{3}\left( \frac{4\pi \alpha}{x_{W}}\right)^{2}(D_{L}\Sigma F^{2}_{L}+D_{R}\Sigma F^{2}_{R}) \\
     & - & \frac{16}{5}\lambda^{2}\left( \frac{E}{M_{H}}\right)^{8}(D_{L}+D_{R}) \\
c_{1} & = & \left( \frac{4\pi \alpha}{x_{W}}\right)^{2}(D_{L}\Sigma F^{2}_{L}+D_{R}\Sigma F^{2}_{R}) \\
     & + & 4\lambda \left( \frac{4\pi \alpha}{x_{W}}\right)\left( \frac{E}{M_{H}}\right)^4(D_{L}(F_{LL}-F_{LR})+D_{R}(F_{RR}-F_{RL})) \\
     & + & 8\lambda^{2}\left( \frac{E}{M_{H}}\right)^{8}(D_{L}+D_{R})  \\
c_{3} & = & \left( \frac{4\pi \alpha}{x_{W}}\right)^{2}\frac{1}{3}(D_{L}\Sigma F^{2}_{L}+D_{R}\Sigma F^{2}_{R}) \\
     & - & 4\lambda \left( \frac{4\pi \alpha}{x_{W}}\right)\left( \frac{E}{M_{H}}\right)^4(D_{L}(F_{LL}-F_{LR})+D_{R}(F_{RR}-F_{RL})) \\
     & - & 8\lambda^{2}\left( \frac{E}{M_{H}}\right)^{8}(D_{L}+D_{R})  \\
c_{5} & = & \frac{32}{5}\lambda^{2}\left( \frac{E}{M_{H}}\right)^{8}(D_{L}+D_{R}) \\
c_{2},c_{4} & = & 0 \\
d_{CE}& = & \left( \frac{4\pi \alpha}{x_{W}}\right)^{2}\frac{2}{3}(D_{L}\Sigma F^{2}_{L}+D_{R}\Sigma F^{2}_{R})+\frac{16}{5}\lambda^{2}\left( \frac{E}{M_{H}}\right)^{8}(D_{L}+D_{R})
\end{eqnarray*}
\subsection{\underline{SR}}
\begin{eqnarray*}
c_{0} & = & -\frac{2}{3}(D_{L}\Sigma F^{2}_{L}+D_{R}\Sigma F^{2}_{R}) \\
     & - & \frac{\pi^{2}}{6}\left( \frac{E}{M_{S}}\right)^{4}\left( D_{L}(\Sigma F^{2}_{L}+3T\Sigma F_{L})+D_{R}(\Sigma F^{2}_{R}+3T\Sigma F_{R}) \right) \\
     & - & \frac{\pi^{4}}{90}\left( \frac{E}{M_{S}}\right)^{8}\left( D_{L}(\Sigma F^{2}_{L}+6T\Sigma F_{L})+D_{R}(\Sigma F^{2}_{R}+6T\Sigma F_{R})+18T^{2}(D_{L}+D_{R}) \right) \\
c_{1} & = & (D_{L}\Sigma F^{2}_{L}+D_{R}\Sigma F^{2}_{R}) \\
     & + & \frac{\pi^{2}}{6}\left( \frac{E}{M_{S}}\right)^{4}\left( D_{L}(\Sigma F^{2}_{L}+3T\Sigma F_{L})+D_{R}(\Sigma F^{2}_{R}+3T\Sigma F_{R}) \right) \\
     & + & \frac{\pi^{4}}{144}\left( \frac{E}{M_{S}}\right)^{8}\left( D_{L}(\Sigma F^{2}_{L}+6T\Sigma F_{L})+D_{R}(\Sigma F^{2}_{R}+6T\Sigma F_{R})+18T^{2}(D_{L}+D_{R}) \right) \\
c_{3} & = & \frac{1}{3}(D_{L}\Sigma F^{2}_{L}+D_{R}\Sigma F^{2}_{R}) \\
     & + & \frac{\pi^{2}}{6}\left( \frac{E}{M_{S}}\right)^{4}\left( D_{L}(\Sigma F^{2}_{L}+3T\Sigma F_{L})+D_{R}(\Sigma F^{2}_{R}+3T\Sigma F_{R}) \right) \\
     & + & \frac{\pi^{4}}{72}\left( \frac{E}{M_{S}}\right)^{8}\left( D_{L}(\Sigma F^{2}_{L}+6T\Sigma F_{L})+D_{R}(\Sigma F^{2}_{R}+6T\Sigma F_{R})+18T^{2}(D_{L}+D_{R}) \right) \\
c_{5} & = & \frac{\pi^{4}}{720}\left( \frac{E}{M_{S}}\right)^{8}\left( D_{L}(\Sigma F^{2}_{L}+6T\Sigma F_{L})+D_{R}(\Sigma F^{2}_{R}+6T\Sigma F_{R})+18T^{2}(D_{L}+D_{R}) \right) \\
c_{2},c_{4} & = & 0 \\
d_{CE}& = & \frac{2}{3}(D_{L}\Sigma F^{2}_{L}+D_{R}\Sigma F^{2}_{R}) \\
     & + & \frac{\pi^{2}}{6}\left( \frac{E}{M_{S}}\right)^{4}\left( D_{L}(\Sigma F^{2}_{L}+3T\Sigma F_{L})+D_{R}(\Sigma F^{2}_{R}+3T\Sigma F_{R}) \right) \\
     & + & \frac{\pi^{4}}{90}\left( \frac{E}{M_{S}}\right)^{8}\left( D_{L}(\Sigma F^{2}_{L}+6T\Sigma F_{L})+D_{R}(\Sigma F^{2}_{R}+6T\Sigma F_{R})+18T^{2}(D_{L}+D_{R}) \right) \\
\end{eqnarray*}

\end{document}